\documentclass[pre,showpacs]{revtex4}
\usepackage{graphicx}
\usepackage{bm}

\begin{document}
\topskip 20mm

\title{ Criticality in the Quantum Kicked Rotor with a Smooth Potential}
\author{Rina Dutta and Pragya Shukla}
\affiliation{Department of Physics, Indian Institute of Technology,
Kharagpur, India.}
\date{\today}
 \widetext

 \begin{abstract}

    We investigate the possibility of an Anderson
   type transition in the quantum kicked rotor with a smooth potential due
   to dynamical localization of the wavefunctions.
   Our results show the typical characteristics of a critical behavior i.e
   multifractal eigenfunctions and
   a scale-invariant level-statistics at a critical kicking
   strength which classically corresponds to a mixed regime.
   This indicates the existence of a localization
   to delocalization transition in the quantum kicked rotor.

       Our study also reveals the possibility of other type of transitions
   in the quantum kicked rotor, with a kicking strength well within strongly
   chaotic regime. These transitions, driven by the breaking of exact
   symmetries e.g. time-reversal and parity, are similar to
   weak-localization transitions in disordered metals.

 \end{abstract}



 \maketitle

\section {Introduction}
 .

   The analogy of the statistical fluctuations of dynamical systems and
disordered systems is  well-known in the delocalized wave-regime
(corresponding to the metallic limit in disordered systems and the
classically chaotic limit in dynamical systems) and has been
explained using random matrix theory as a tool \cite{casati1, izrp,
fyd, thaha}. A similar analogy exists for fully localized regimes of
the wavefunctions too (i.e. between the insulator limit of
disordered systems and the integrable limit of dynamical systems)
\cite{casati1, izrp, fyd}. It is therefore natural to probe the
presence/ absence of the analogy in partially localized/ critical
regimes of these systems.  Our analysis shows that, similar to the 
$d>2$ Anderson Hamiltonian ($d$ as dimension), the $d=1$ quantum 
kicked rotor (QKR) undergoes a localization-delocalization transition 
in the classically mixed regime. We also find quantum phase transitions in
its chaotic regime due to breaking of the symmetries e.g.
time-reversal and parity in the quantum system. Similar to
disordered systems,  the symmetry-breaking transitions in the QKR
occur due to weak-localization effects. Similar phase transitions
due to symmetry-breakings have been seen in a few other complex
systems too e.g. the ensembles of distinguishable spins \cite{pizo}.

   The connection of the kicked rotor to the $d=1$ Anderson
Hamiltonian  has been known for several decades \cite{fgp, izrp,
casati2, chir}. A recent work \cite{garcia} further explores the
connection and shows that, for the non-analytic potentials in the QKR,
the eigenstates show multifractality or power-law localization
\cite{mj, caste, cu1, emm}, a behavior similar to the eigenstates of
a $d>2$ dimensional Anderson Hamiltonian \cite{leer} at its critical
point. Our study shows existence of the multifractal eigenstates in
the QKR with smooth potentials too e.g. $V(q)=K {\rm cos}(q)$ at
specific parametric conditions. Furthermore, similar to a critical
Anderson system, the multifractality in the QKR is accompanied by a
critical level statistics (size-independent and different from the
two ends of the transition), a necessary criteria for the critical
behavior \cite{mj}. This indicates a much deeper connectivity of the
kicked rotor to the Anderson Hamiltonian, not affected just by the
nature of the potential or the dimension of system. As discussed
here, the connection seems to be mainly governed  by the "degree of
complexity" (measured by the complexity parameter discussed later)
and  may exist among a wider range of dynamical and disordered
systems.

The present study is motivated by a recent analytical work
\cite{psco, psal} leading to a common mathematical formulation for
the statistical fluctuations of a wide range of complex systems. The
work, based on the ensemble-averaging, shows that the fluctuations
are governed by a single parameter $\Lambda$ besides
global-constraints on the system \cite{psco, psal}. $\Lambda$
referred as the complexity parameter, turns out to be a function of
the average accuracy of the matrix elements, measured in units of
the mean-level spacing. The fluctuations in two different systems,
subjected to similar global-constraints, are analogous if their
complexity parameters are equal irrespective of other
system-details.

The $\Lambda$-formulation was recently used by us to find the
Gaussian Brownian ensemble (GBE) \cite{me} analog and the power-law
random banded matrix ensemble (PRBME) \cite{mir} analog of the 
Anderson system (for arbitrary $d$) \cite{psand}.
However it can not directly be applied to find the QKR analog; this
is, in principle, due to inapplicability of the ensemble averaging
to dynamical systems. Fortunately it is possible to derive $\Lambda$
for dynamical systems by a semi-classical route, using the
phase-space averages \cite{psnl}. The "semi-classical" $\Lambda$ was
used by us \cite{psnl,psef} to map the statistics of the
time-evolution operator $U$ of the QKR to the circular Brownian
ensembles (CBE) \cite{me}. The lack of a suitable criteria for the
critical statistical behavior prevented us earlier from a critical
QKR-analysis. Our present work pursues the analysis by first
analytically identifying the critical QKR-behavior using the
"semi-classical" $\Lambda \equiv \Lambda_{kr}$; the limit $N
\rightarrow \infty \; \Lambda \rightarrow \Lambda^*$, with
$\Lambda^*$ independent of the size $N$, gives the critical points
of transition. This is followed by a numerical analysis of an
ensemble of the QKR which confirms critical behavior at the
semi-classically predicted values. This in turn suggests a
paradoxical validity of the ensemble averaging in dynamical systems
at least of QKR type; (also indicated by the CBE-QKR mapping). A
subsequent comparison of the "semi-classical" $\Lambda_{kr}$ to the
"ensemble-based" $\Lambda_{ae}$ of the Anderson
Hamiltonian (arbitrary $d$) gives us its QKR analog; the analogy is 
numerically confirmed too.

The statistical behavior in the bulk of the spectrum of a standard
Gaussian ensemble is known to be analogous to a standard circular
ensemble (for large matrix sizes) \cite{me, apps}. Our work extends
this analogy to their Brownian ensemble counterparts too, that is,
between the GBE and the CBE (by the mapping GBE $\rightarrow$ AE
$\rightarrow$ QKR $\rightarrow$ CBE). This in turn indicates a
connection among a wide-range of physical systems which are known to
be well-modeled by the Anderson Hamiltonian, the kicked rotor and
the  Brownian ensembles of Gaussian and circular type.

The paper is organized as follows: The section II briefly reviews
the basic features of the kicked rotor and the Anderson Hamiltonian
required for our analysis; it also discusses the parametric
conditions in the kicked rotor which can support critical points.
The section III deals with the numerical confirmation of the
critical level-statistics and the multifractality of the
eigenfunctions at critical parametric conditions in the QKR  and
their comparison with a $d=3$ dimensional Anderson Hamiltonian. We
conclude in section IV with a brief discussion of our main results
and open questions.

\section {Kicked Rotor and Anderson Hamiltonian}

        The kicked rotor and the Anderson Hamiltonian have been subjects of intense
study in past and many of their details can be found in several
references \cite{casati1, izrp, casati2, been, thaha}. However, for
self consistency of the paper, we present here a few details
required for later discussion.


\subsection {Kicked Rotor and Complexity Parameter}

The kicked rotor can be described as a pendulum subjected to
periodic kicks (of time-period $T$) with Hamiltonian $H$ given as

\begin{eqnarray}
H={(p+\gamma)^2\over 2 } + K \; {\rm cos}(\theta +\theta_0)
\sum_{n=-\infty}^{\infty} \delta(t-n T). \label{H}
\end{eqnarray}
Here $K$ is the stochasticity parameter, $\gamma$ and $\theta_0$ are
the time-reversal and parity symmetry breaking parameters in the
quantum Hamiltonian while acting in a finite Hilbert space.

	Integration of the equations of motion 
$\dot \theta = -{\partial H \over \partial p}$,  
$\dot p = {\partial H \over \partial \theta}$ between subsequent 
kicks e.g. $n$ and $n+1$ gives the classical map, 

\begin{eqnarray}
p_{n+1} &=& p_n + K \; {\rm sin}(\theta_n +\theta_0) \qquad \qquad 
({\rm mod}\; 2\pi), \nonumber \\
\theta_{n+1} &=& \theta_n + p_{n+1} \qquad \qquad 
({\rm mod}\; 2\pi)
\label{map}
\end{eqnarray}
   The map is area-preserving and invariant under the discrete 
translation $\theta \rightarrow \theta + 2\pi, p \rightarrow p + 2\pi$. 
It also preserves the time-reversal symmetry $p \rightarrow 2\pi-p, 
\theta \rightarrow \theta, t \rightarrow -t$ and the parity 
 $p \rightarrow 2\pi-p, \theta \rightarrow 2\pi-\theta$ for all 
values of $\gamma$ and $\theta_0$.
Thus the classical dynamics depends only on $K$, changing 
from integrable ($K=0$) to near integrable ($0<K<4.5$) to large scale 
chaos ($K>4.5$).

  The quantum dynamics can be described by a
  discrete time-evolution operator $U=G^{1/2}.B.G^{1/2}$ \cite{izrp, chir} 
where
\begin{eqnarray}
B={\rm exp}\left[-i k {\rm cos}(\theta+\theta_0) \right] \\
G={\rm exp}\left[-i T (p+\gamma)^2/2\hbar \right] ={\rm exp}\left[i
\tau ({\partial \over \partial \theta} - i {\gamma\over \hbar})^2/2
\right] \label{BG}
\end{eqnarray}
Here $k=K/\hbar$, $\tau=T\hbar$, $\theta$ and $p=i\hbar
{\partial\over \partial \theta}$ are the position and momentum
operators respectively; $p$ has discrete eigenvalues,
$p|m>=m\hbar|m>$ ($m=1 \rightarrow N$) due to periodicity of
$\theta$, ($\theta \rightarrow \theta+2\pi$). The choice of a
rational value for $\tau/4\pi=M/N$ results in a periodicity also for
momentum operator $p'=p+4\pi M/T$, ($l'=l+N$) and therefore in
discrete eigenvalues for $\theta$, ($\theta |l>=(2\pi l/N)|n>$). The
quantum dynamics can then be confined to a two dimensional torus
(with a Hilbert space of finite size $N=2\pi m_0/\tau$ with $m_0$ an
integer). The classical analog of this model corresponds to the
standard mapping on a torus of size $2\pi m_0/T$ in the momentum
$p$; thus the classical limit is  $\tau \rightarrow 0, k\rightarrow
\infty, N\rightarrow \infty$ with $K={\rm constant}$ and $N\tau
={\rm constant}$ \cite{chir, izrp}.

    For the dynamical-localization analysis, it is useful to express
the matrix $U$ in the momentum-basis \cite{izrp, izrp, psnl}:

\begin{eqnarray}
U_{mn}={1\over N} {\rm exp}\left[i \tau (m -\gamma/\hbar)^2/4 +
i\tau (n -\gamma/\hbar)^2/ 4 \right] \sum_{l=-N_1}^{N_1}  {\rm
exp}\left[-i\; k \; {\rm cos}\left(2\pi l /N + \theta_0
\right)\right] {\rm exp}\left[-2 \pi i l (m-n)/N \right] \label{U}
\end{eqnarray}
where $n,m=-N_1,...,N_1$ with $N_1=(N-1)/2$  if $N$ is odd and $N_1=N/2$ if $N$
is even. It is clear that the properties of $H$, eq.(\ref{H}), are
recovered in the infinite matrix size limit.

The quantum dynamics under exact symmetry conditions ($\gamma=0$,
$\theta_0=0$) can significantly be affected by relative values of
$k,\tau, N$. It was first conjectured  \cite{izrp} and later on
verified \cite {psnl} that the statistical properties are governed
by the ratio of the localization length $\zeta$ to the total number
of states $N$ or, equivalently, $k^2/ N$ (as $\zeta=D/2\tau^2$ with
$D \approx K^2/2$ as the diffusion constant). However, to best of
our knowledge, the critical behavior at $k^2 \simeq N$ was not
probed before. The other parameters playing a crucial role in the
quantum dynamics are $\gamma$ and $\theta_0$, the measures of
time-reversal and parity-symmetry breaking respectively (with $0 <
\gamma <\hbar$ and $-\pi/N < \theta_0 < \pi/N$). Note the change of
$p \rightarrow p+\gamma$ or $\theta \rightarrow \theta + \theta_0$
is a canonical transformation, thus leaving the classical
Hamiltonian unaffected. The corresponding quantum dynamics, however,
is affected as the quantum Hamiltonian acting in a finite Hilbert
space may not remain invariant under a unitary transformation.

Following eq.(\ref{U}), the multi-parametric nature of  $U$ is
expected to manifest itself in the statistical behavior of its
eigenvalues (quasi-energies) and the eigenfunctions. However, as
shown in \cite{psnl} using semi-classical techniques, the
quasi-energy statistics of $U$ is sensitive to a single parameter
$\Lambda_{kr}$ and the exact symmetry-conditions. Under exact
time-reversal symmetry ($\gamma=0$) and partially violated parity
($\theta_0 \not=0$) (taking $T=1$, equivalently, $\tau=\hbar$, 
without loss of generality), we have \cite{psnl}
\begin{eqnarray}
\Lambda_{\rm kr,t}= {\theta_0^2 N k^2 \over 4 \pi^2 } = 
{N^3 \theta_0^2  K^2 \over 64 \pi^4 M^2 } \label{lamkrt}
\end{eqnarray}
Note, for $\theta_0=\pi/2N$, $\Lambda_{\rm kr, t}$ is essentially the same
as the one conjectured in \cite{izrp} for scaling behavior of the
spectral statistics.
Similarly, for the strongly chaotic case ($k^2 > N$) with only
parity symmetry ($\theta_0=\pi/2N)$ and no time-reversal
($\gamma\not=0$),
\begin{eqnarray}
\Lambda_{\rm kr,nt}= { \gamma_q^2 N^3 \over 48 \pi^2 } 
\label{lamkrnt}
\end{eqnarray}
with $\gamma_q=\gamma/\hbar$. 
As eq.(\ref{lamkrt}) indicates, $\Lambda_{kr,t} \rightarrow \infty$
in the strongly chaotic limit $k \rightarrow \infty$; similarly,
from eq.( \ref{lamkrnt}), $\Lambda_{kr,t} \rightarrow \infty$ for
$\gamma \simeq \hbar$; the statistics in these cases can be
well-modeled \cite{iz1, psnl} by the standard random matrix
ensembles of unitary matrices, known as the standard circular 
ensembles e.g. circular orthogonal ensemble (COE), circular 
unitary ensemble (CUE)
etc\cite{me}. The cases with a slow variation of $k$, $\gamma$ or
$\theta_0$ (partial localization, partial time-reversal or
parity-violation respectively) and finite size $N$ correspond to a
smooth variation of $\Lambda_{kr,t}$ or $\Lambda_{kr,nt}$ between
$0$ and $\infty$. The intermediate statistics for these cases
\cite{psnl, psef} can be well-described by the circular Brownian
ensembles. The latter are the ensemble of unitary matrices,
described as $U_{w}=U_0^{1/2}{\rm exp}[i w V] U_0^{1/2}$ and
characterized by a single parameter $\Lambda_{cbe}= w^2 \langle |V_{
kl}|^2 \rangle /D^2$ ($D=2\pi/N$) and the exact system-symmetries 
\cite{me,apps}. The perturbation $V$ belongs to a standard Gaussian 
ensemble
of the Hermitian matrices e.g Gaussian orthogonal ensemble (GOE),
Gaussian unitary ensemble (GUE) etc \cite{me}. The circular Brownian
ensemble analog of a QKR is given by the condition
\begin{eqnarray}
\Lambda_{kr}=\Lambda_{cbe} . \label{krbe}
\end{eqnarray}
As mentioned above, the studies \cite{psnl, psef} did not explore
the infinite size limit of $\Lambda_{kr}$ and its application for
the critical behavior analysis; we discuss it in next section.

\subsection{Critical behavior of Quantum Kicked Rotor}

For a critical point analysis of the QKR, we search for  the system
conditions leading to the critical eigenvalue-statistics and the
multifractal eigenfunctions in the infinite size limit ($N \rightarrow
\infty$). The parametric conditions for the critical level
statistics, characterized by a non-zero, finite $\Lambda$ in the
limit $N\rightarrow \infty$, can be obtained from eqs.(\ref{lamkrt},
\ref{lamkrnt}). The analysis suggests the possibility of several
continuous families of critical points, characterized by the
complexity parameter and  the exact symmetries; we mention here only
the three main cases:

(i) $k \propto \sqrt{N}, \gamma=0, \theta_0=\pi/2N$:
Eq.(\ref{lamkrt}) in this case leads to a size-independent
$\Lambda_{kr,t}=\chi^2/16$ with $\chi=k/\sqrt{N}$. For small-$k$ (in
the mixed regime), this corresponds to a localization $\rightarrow$
delocalization transition under the time-reversal conditions (no parity
symmetry) with a continuous family of critical points characterized
by $\chi$. The bulk-statistics here is analogous to a circular
Brownian ensemble $U_w$ (see eq.(\ref{krbe})) with $w=\chi \pi/2
\sqrt { N}$ (due to a GOE type perturbation $V$ with
$|V_{kl}|^2=(1+\delta_{kl})$ of a Poisson matrix \cite{me}, see
eq.(\ref{krbe})). The two ends of the transition in this case are
the Poisson ($\Lambda \rightarrow 0$) and the COE ensemble ($\Lambda
\rightarrow \infty$).

(ii) $ \gamma_q \propto  N^{-3/2}, \theta_0=\pi/2N$: Eq.(\ref{lamkrnt})
in this case gives $\Lambda_{kr,nt}=\lambda^2/ 48 \pi^2$ with
$\lambda=\gamma_q N^{3/2}$. For large $k$ (in strongly chaotic limit),
a finite $\lambda$ gives the critical parameter for the transition
from a time-reversible  to a
time-irreversible phase (both phases delocalized); it can be referred as the
weak-localization critical point. The end-points of the transition
are the COE ($\Lambda \rightarrow 0$) and the CUE ($\Lambda \rightarrow
\infty$), with the critical statistics given by an intermediate circular
Brownian ensemble with $w = \lambda /(2 N\sqrt{3})$ (due to a GUE
type perturbation $V$,  with $|V_{kl}|^2=1$, of a COE matrix
{\cite{me} ).

(iii) $ \gamma=0, \theta_0  \propto N^{-3/2}$: Here
eq.(\ref{lamkrt}) gives $\Lambda_{kr,t}= \phi^2 K^2/(64 \pi^4 M^2)$ with
$\phi= N^{3/2} \theta_0$ describing  the critical point family for
the transition from a parity-symmetric phase to
a parity fully violated phase (both phases time-reversal). 
For $K$ in the mixed regime ($K < N\hbar$), a variation of $\phi$ leads to 
the Poisson $\rightarrow$ COE transition.  
For $K$ in the strongly chaotic regime ($K > N\hbar$), the transition
end-points are the 2-COE ($\Lambda \rightarrow 0$) and the COE
($\Lambda \rightarrow \infty$). The critical statistics for a specific $K$ 
is given by the intermediate Brownian ensemble with $w=K \phi /(4\pi M N)$
(due to a GOE type perturbation $V$, with $|V_{kl}|^2=1$, of the Poisson 
ensemble if $K << N\hbar$, or, the 2-COE
ensemble if $K>>N\hbar$).

As eqs.(\ref{lamkrt}, \ref{lamkrnt}) indicate, a size-independent
$\Lambda_{kr}$ can be obtained by other combinations of $k, \theta,
\gamma$ too. This suggests critical behavior in the symmetry-spaces
other than those mentioned above.

The critical nature of the system for specific parametric conditions
can further be confirmed by an analysis of the eigenfunction
fluctuations. The studies of a wide range of systems (see \cite{mj,
caste, cu1, emm} and the references therein) reveal the presence of
strong fluctuations in the eigenfunctions near a critical point. The
fluctuations can be characterized through the set of generalized
fractal dimension $D_q$ or $\tau_q=(q-1)D_q$, related to the scaling of
the $q^{\rm th}$ moment of the wavefunction intensity $|\phi(r)|^2$
with size $N$ \cite{caste}: $P_q = \int {\rm d}r |\phi(r)|^{2q} =
N^{-\tau(q)/d}$. The multifractality of the eigenfunctions can also
be analyzed through the spectrum of singularity strengths
$f(\alpha)$ \cite{mj}, related to $\tau(q)$ by a Legendre
transformation: $f(\alpha(q))=q\alpha(q) -\tau(q)$ (see
\cite{mj,emm} for details). In section III, we numerically analyze
both $\tau(q)$ and $f(\alpha)$ to detect the multifractality of the
QKR-eigenfunctions.

\subsection{Anderson Hamiltonian and the Complexity Parameter}

The Anderson model for a disordered system is described by
 a $d$-dimensional disordered lattice, of size $L$, with a
  Hamiltonian $H= \sum_n \epsilon_n a_n^+ a_n -
\sum_{n\not=m} b_{mn} (a_n^+ a_m +a_n a_m^+)$ in the tight-binding
approximation \cite{leer}. In the site representation, $H$ turns out
to be a sparse matrix of size $N=L^d$ with the diagonal elements as
the site energies $H_{kk}=\epsilon_k$ and the off-diagonals
$H_{mn}=b_{mn}$ given by the hopping conditions. For a Gaussian type
on-site disorder (of variance $\omega$ and zero mean) and a nearest
neighbor (n.n.) isotropic hopping with both random (Gaussian) and/
or non-random components, $H$ can be modeled by an ensemble (later
referred as the Anderson ensemble or AE) with following density

 \begin{eqnarray}
 \rho (H,v,b)=C {\rm exp}\left[- \sum_k H_{kk}^2/2 \omega
- \sum_{(k,l)=n.n.} H_{kl}^2/2 \eta \right]
\prod_{(k,l)=n.n} \delta (H_{kl}-t)
\prod_{(k,l) \not=n.n.} \delta (H_{kl})
\label{eq5}
\end{eqnarray}
with $C$ as the normalization constant.
As discussed in \cite{psand}, the above ensemble can be rewritten as
\begin{eqnarray}
\rho (H,h,b)=C{\rm exp}[{- \sum_{k\le l} (1/2 h_{kl})
(H_{kl}-b_{kl})^2 }]
\end{eqnarray}
where $b_{kl}=0$, $h_{kk} = \omega, h_{kl} = \eta f_{kl}$
with $f(kl)=1$ for $\{k,l\}$ pairs connected by the hopping,
$f(kl)\rightarrow 0$ for all $\{k,l\}$ pairs representing
the disconnected sites.
The single parameter $\Lambda_{\rm ae}$
governing the spectral statistics (see eq.(19) of \cite{psand}) can then
be given as
\begin{eqnarray}
\Lambda_{a,e}(E,Y) 
&=& \left(|\alpha_w-\alpha_i| F^2\over  \gamma_0\right)
\zeta^{2d} L^{-d}
\approx {|\alpha_w-{\alpha_i}|\over {\gamma_0} N}
{\left(F\over  I_2^{typ}\right)}^2,
\label{lamae}
\end{eqnarray}
with
\begin{eqnarray}
 \alpha_w={\rm ln}|1-\gamma_0 \omega| +
(z/2) {\rm ln}\left[|1- 2\gamma_0  \eta||t+\delta_{t0}| \right]
\label{alpha}
\end{eqnarray}
 with  $z$ as the number of the nearest-neighbors, 
$\alpha_i= -{\rm ln} 2$ and $\gamma_0$
 as an arbitrary constant.
Further $F(E)$ is the mean level density, $\zeta$ as the
localization length and  $I_2^{typ}$ as the typical inverse
participation ratio: $I_2^{typ} \propto \zeta^{-d}$.

\section {numerical Analysis}

        The objectives of our numerical analysis are two fold:
(i) a search for the critical points of the quantum kicked rotor,
and, (ii) a comparison of its fluctuation measures with those of a
3-dimensional Anderson ensemble. For this purpose, we analyze the
following cases:

\vspace{.2in}

(i) {\bf QKR1: quantum dynamics time-reversal but parity broken}: 
$k^2=\chi^2 N, \chi \approx 1.5, \gamma=0, \theta_0=\pi/2N, 
T=1, \tau=\hbar =40 \pi/N$ which gives $K
\approx 189 /\sqrt{N}$}. This case corresponds to 
the critical set (i) in section II. B and is  
analyzed for many sizes ($N=213 \rightarrow 1013$) to verify 
the critical behavior.  

\vspace{.2in}

(ii) {\bf QKR2: quantum dynamics with both time-reversal and 
parity broken}:  $k \approx 20000, \gamma= \lambda_q N^{-3/2}, 
\lambda_q=6, T=1, \theta_0=\pi/2N, \tau=\hbar=40\pi/N$. 
This case belongs to the set (ii) in section II.B and its 
critical nature is also confirmed by analyzing many $N$ 
values. 

\vspace{.2in}

(iii) {\bf QKR3}: : same as QKR1 but with $\chi=0.8$ which gives 
$K \approx 100 /\sqrt{N}$. We consider this case to verify the 
analogy with a $d=3$ Anderson ensemble.

\vspace{.2in}

(iv) {\bf QKR4: quantum dynamics time-reversal but parity broken}: 
$k \approx 4.5 /\hbar, \gamma=0, \theta_0=\phi N^{-3/2}, \phi=0.84 \pi^2, 
T=1,\tau=\hbar =8 \pi/N $ (case (iii) in section II.B). 
This is also analogous to the Anderson system mentioned above in 
the QKR3 case, notwithstanding the crucial changes in 
$k$ and $\theta_0$ for the two QKR cases.

To explore critical behavior, we analyze large ensembles of
the matrices $U$ for both QKR1 and QKR2 for various matrix sizes
$N$; the ensemble in each case is obtained by varying $k$ in a small
neighborhood while keeping $N$ fixed. The chosen $N$ range give $K$
in the mixed regime for QKR1 ($6.25 <K < 13$) and in the chaotic regime
for QKR2 ($1000 < K < 12000$). Prior to the analysis, the quasi-energies
(the eigenvalues of $U$) are unfolded by the local mean level
density $D^{-1}$ ($=N/2\pi$, a constant due to repulsion and 
a unit-circle confinement of the quasi-energies). 
The figures 1,2 display the nearest-neighbor spacing distribution 
$P(s)$ and the number variance $\Sigma^2(r)$, the measures of the 
short and long-range spectral
correlations respectively, for the QKR1 and the QKR2. Note the curves in
figure 1 are intermediate to the Poisson and the COE limits; the
size independence implies their survival
 in the infinite size limit too. This  indicates the 
QKR1 as the critical point of transition from a localized phase to
the delocalized phase. Similarly the curves in figure 2,
intermediate to the COE and the CUE limits, suggest the QKR2 as the
critical point of transition from the time-reversed phase to the
time-irreversible phase (both phases in the chaotic regime).

To reconfirm the critical nature of the QKR1 and the QKR2, we
numerical analyze the moments of their local eigenfunctions
intensity for various sizes. The results shown for $\tau_q$ in
figures 3a,4a indicate the multifractal nature of the
eigenfunctions; for small-$q$ ranges, $\tau_q$ shows a behavior
$\tau_q=(1+c)\;q-d-c\;q^2$ (or $D_q=1-c q$) with $d=1$ and $c=0.06,
0.075$ for the QKR1 and the QKR2 respectively (see table 1 for the first few
values of $D_q$). These results are reconfirmed by a numerical study
of the $f(\alpha)$ spectrum displayed in figures 3b,4b. For this
purpose, we use the procedure based on the evaluation of moments,
described in \cite{emm} (using eq.(4) and eq.(10) of \cite{emm})
which has the advantage of full control over the finite-size
corrections. The numerical results for $\alpha_0, \alpha_1$ and
$\alpha_{1/2}$ for the QKR (see table 1) indicate a parabolic form of
$f(\alpha)$ (also confirmed by the fits shown in figure 3b, 4b) and
therefore a log-normal behavior of the local eigenfunction intensity
$u=|\psi(r)|^2$ for large $u$-regions (see \cite{emm, mj, caste} for
details). As shown in figures 3c, 4c, the tail-behavior of $P_u(u')$
can be well-approximated by the function
\begin{eqnarray}
f(u')=\sqrt{a/2\pi} {\rm e}^{b+a u'} {\rm e}^{-a( u'+c)^2/2)}.
\label{ln}
\end{eqnarray}
with $u'=[{\rm ln}u-\langle {\rm u} \rangle]/{\langle {\rm ln}^2 u \rangle}$
(see figures 3c, 4c for numerical values of a,b,c).

The bulk statistical behavior of the Hermitian matrices is known to
be analogous to the unitary matrices \cite{me, apps}. This, along
with the single parametric formulation of the statistics of the
Hermitian matrix ensemble \cite{psco}, suggests the analogy of the
QKR-Anderson ensemble statistics if their
$\Lambda$ parameters are equal (besides similar global constraints
e.g. global symmetries) \cite{psal}. Our next step is the comparison
of the fluctuation measures of a time-reversal Anderson case  with
the QKR3, a time-reversal system with partially localized wavefunction
in the momentum space (dynamical localization). For this purpose, we
analyze a cubic ($d=3$) Anderson lattice of linear size $L$
($N=L^d$) with a Gaussian site disorder (of variance
$\omega=W^2/12$, $W=4.05$ and mean zero), same for each site,
 an isotropic Gaussian hopping (of variance $\eta=1/12$ and mean zero)
between the nearest-neighbors
 with hard wall boundary conditions; these condition
correspond to the critical point for a disorder driven metal-insulator
transition \cite{psand}). A substitution of the above values (with
$t=0$) in eq.(\ref{alpha}) gives
$\alpha_w-\alpha_i=1.36$. As shown by the numerical
analysis in \cite{psand, pswf}, $F(E) \approx 0.26 e^{-E^2/5}$ and
$I_2^{typ} \approx 0.04$ which on substitution in eq.(\ref{lamae}) gives
$\Lambda_{ae}=0.056$ (with $\gamma_0=2$).

For AE-QKR comparison, we analyze the ensembles of 2000 matrices with
matrix size $N=L^3=512$ for the 3d AE case and $N=513$ for the QKR
case. The energy dependence of $\Lambda$ (see \cite{psand, psnl}) 
forces us to confine our analysis to only $10{\rm \%}$
of the levels near the band-center from each such matrix. The levels
are unfolded by respective local mean-level density in each case (so
as to compare the level-density fluctuations on a same
density-scale) \cite{me}. Figure 5a shows the AE-QKR3-QKR4
comparison of $P(s)$; the good agreement among the three curves 
verifies the $\Lambda$-dependence of the spectral correlations.
This is reconfirmed by figure 5b showing a comparison of the
number variance. Note that $\phi \approx 0.84 \pi^2$ for QKR4 
 is same as the theoretical analog given by the condition 
$\Lambda_{kr,t}=\Lambda_{ae}=0.056$ 
(see eqs.(\ref{lamkrt},\ref{lamae})). However $\chi=0.8$ for 
QKR3 show a small deviation from the  theoretically prediction 
($\chi \approx 0.95$).  
This may be due to eq.(\ref{lamkrt}) being a poor approximation 
at the integrable-nonintegrable boundary $K=0$. 
Note that the classical limits of QKR3 and QKR4 are different 
($K=0$ for QKR3, $K=4.5$ for QKR4),  
although their "semi-classical" $K$ are same
($K=4.5$ with $N=513$ for QKR3, $K=4.5$ for any $N$ for QKR4). 
As clear from eq.(\ref{map}),  
$K=0$ marks the boundary between the integrable and mixed dynamics; 
$K=4.5$ corresponds to the mixed nature of the dynamics (see figure 5c).

        As discussed in \cite{pswf}, the eigenfunction fluctuations
of  finite systems are influenced by two parameters, namely, system 
size $N$ as well as $\Lambda_{measure}$. To compare
$\Lambda_{measure}$ dependence of an eigenfunction measure,
therefore, same system size should be taken for each system.
Figure 6a shows the distribution  $P_u (u')$ of the local 
eigenfunction intensity, 
$u'=[{\rm ln}u-\langle {\rm ln}u \rangle]/ \langle {\rm ln}^2 u \rangle$,
for the ${\rm AE}$ and the QKR3. 
The close proximity of the two curves suggests $\Lambda_{kr,t}$ 
as the parameter governing the local eigenfunctions intensity too;  
(we have verified the analogy also with QKR4).  
A comparison of the AE-QKR3 multifractality spectrum, 
shown in figure 6b, reconfirms their close similarity  
at least on the statistical grounds. 

      The  numerical confirmation of the   
statistical analogy of QKR4-AE-QKR3 systems supports 
our claim regarding single parametric ($\Lambda$)-dependence of the 
statistics besides global constraints. Note the latter are same for 
both QKR3 and QKR4 (i.e parity violated, time-reversal preserved and 
mixed dynamics) which results in their statistics intermediate to  
same universality classes, namely, Poisson and COE although the 
transition parameters are different in the two cases.

         The QKR-AE analogy can be utilized to connect them to other
complex systems too. In \cite{pswf, psand}, we studied the
AE-connection with the  PRBME (described by $\langle H_{kl}
\rangle=0$, $ \langle H_{kl}^2 \rangle  \propto  [1+|k-l|^2/p^2]^{-1}$
\cite{mir}) and the GBE ($\langle H_{kl} \rangle=0$, $ \langle
H_{kl}^2 \rangle \propto [1+c N^2]^{-1}$ \cite{psand}). For the AE
case considered above, the PRBME and the GBE analogs for the spectral
statistics turned out to be $p=0.4$ and $c=0.1$; these systems are
therefore the spectral statistical analogs of  QKR3, QKR4 as well 
as of a $N\times N$ circular Brownian ensemble with $w \approx 
0.4 \pi N^{-1/2}$ (see case(i) of section II.B).

\section{conclusion}

To summarize, we studied the statistical analogy of two paradigmatic
models of dynamical and disordered systems, namely, the quantum
kicked rotor and the Anderson Hamiltonian, in the partially
localized regimes. Our results indicate the existence of critical
behavior in the classically mixed regime of the QKR  
with a smooth potential. This is qualitatively analogous to a disorder
driven metal-insulator transition in the Anderson system; the
quantitative analogy for their statistical behavior follows if their
complexity parameters are equal. Our study also reveals the
possibility of other transitions in the QKR e.g from a symmetry
preserving phase to a symmetry fully violated phase. These
transitions are analogous to similar symmetry breaking transitions
in disordered metals e.g the Anderson Hamiltonian in the weak
disorder limit in the presence of a slowly varying magnetic field.

 As with the Anderson transition, the QKR transitions are governed
 by the complexity parameter $\Lambda$ too.  However, contrary to
the $\Lambda$-derivation for the Anderson case by an ensemble route,
$\Lambda$ for the QKR is derived by a semi-classical method.  The
semi-classical $\Lambda$-formulation is also numerically verified for
the QKR-ensembles. This indicates an equivalence of the
ensemble-averaging and the phase-space averaging for the statistical
analysis. This further lends credence to the single parametric
formulation of the statistical behavior of complex systems,
irrespective of the origin of their complexity. However it needs to
be examined for other dynamical systems.

     Research has indicated a multi-parametric dependence of
     the  spectral-statistics at long energy scales of the dynamical
     systems \cite{berry}, originating in the level-density oscillations
     due to short periodic orbits. However, these studies are not at 
     variance with our work. This is because the "semi-classical" 
     $\Lambda$-derivation in \cite{psnl} is based on the assumed 
     equivalence of the traces of the operators with their phase-space
     averages. The assumption may not be valid on short time-scales of the
     dynamics. One should also understand the exact role of the 
     ensemble-averaging for the statistical analysis of dynamical systems.

\section{acknowledgment} 

We would like to express our gratitude to  Prof. M.V.Berry for many 
helpful suggestions.

\section{Figure Caption}

\noindent Fig. 1 [Color online]. Spectral Measures of QKR1:

(a) the nearest-neighbor spacing distribution $P(S)$ of the
eigenvalues for various sizes $N$, with inset showing behavior on
the linear scale, (b) Number variance $\Sigma^2(r)$ for various
sizes.

The convergence of the curves for different sizes indicates
scale-invariance of the statistics. The behavior is critical, being
different from the two end-points, namely, Poisson and CUE
statistics even in infinite size limit.

\vspace{.5in}

\noindent Fig. 2 [Color online]. Spectral Measures of QKR2: (a)
Distribution $P(S)$ of the nearest-neighbor eigenvalue spacings $S$
for various sizes, with inset showing behavior on the linear scale,
(b) Number variance $\Sigma^2(r)$ for various sizes.

Again the statistics being intermediate between COE and CUE, and,
convergence of the curves for different sizes indicates its critical
behavior.

\vspace{.5in}

\noindent Fig. 3 [Color online]. Multifractality of QKR1 :

(a) Fractal Dimension $\tau_q$ along with the fit $y(q)=(1+c)q-1-c
q^2$ with $c=0.06$ (good only for $q\le 3$). A fit for the large $q$
regime suggest following behavior: $\tau_q=q-1+0.02 q^2$. (b)
Multifractal spectrum $f(\alpha)$  for various sizes along with the
parabolic fit $f(\alpha)=d- {\left(\alpha-\alpha_0 \right)^2 \over 4
\left(\alpha_0-d \right)}$ with $\alpha_0=1.09$ and $d=1$.
(c) Distribution $P_u(u')$  with $u'=[{\rm ln}u-\langle {\rm u}
\rangle]/ {\langle {\rm ln}^2 u \rangle}$  of the local intensity of
an eigenfunction for QKR1. The solid line represent the function
$f(u')$ given by eq.(\ref{ln}) with $a=5.2$, $b=1.2$ and $c=0.78$
(corresponding to an approximate log-normal behavior of $P_u(u)$), a
good approximation in tail-region as expected. The inset shows the
behavior on linear scale.

\vspace{.5in}

\noindent Fig. 4 [Color on line]. Multifractality of QKR2: (a)
Fractal Dimension $\tau_q$ along with the fit $y(q)=(1+c)q-1-c q^2$
with $c=0.075$ (good only for $q<4$). (b) Mulifractal spectrum
$f(\alpha)$ for various sizes along with a parabolic fit, of the
same form as in figure 3.b with $\alpha_0=1.045$ and $d=1$.
(c) Distribution $P_u(u')$  of the local intensity of an
eigenfunction for QKR2, with $u'$ same as in figure 3c. The solid
line represent the function $f(u')$ with $a=5.3$, $b=0.95$ and
$c=0.73$ (corresponding to a log-normal behavior of $P_u(u)$) which
fits well in the tail region of $P_u(u')$ . The inset shows the
behavior on the linear scale.

\vspace{.5in}

\noindent Fig. 5 [Color online]. Comparison of spectral statistics
of the Anderson ensemble with QKR3 and QKR4: (a) $P(S)$, 
with inset showing the linear behavior, 
(b) $\Sigma^2 (r)$. Note  
QKR3 and QKR4  turns out to be close to the one
suggested by the relation $\Lambda_{a}={\Lambda}_{kr,t}$  (giving
$\chi=0.95$ for QKR3 and $\phi=0.84 \pi^2$ for QKR4). 
(c) Phase-space behavior of the classical kicked rotor at $K=4.5$ 
(see eq.(\ref{map}))

\vspace{.5in}

\noindent Fig. 6 [Color online]. Comparison of the multifractality
of the eigenfunctions of QKR3 with the Anderson ensemble: (a) local
intensity distribution $P_u(u')$ of an eigenfunction with $u'$ same
as in figure 3,4. Also shown is the function $f(u')$ with $a=15.9,
b=7.7, c=0.99$, a good fit in the tail region of $P_u(u')$. The
inset shows the behavior on linear scale. (b) multifractal spectrum
$f(\alpha)$: Note the QKR3 analog of the Anderson ensemble here is same
as the one in figure 5. Again the fit has the same parabolic form as
in figures 3,4, with $\alpha_0=1.1$ and $d=1$.

\vspace{1in}

\begin{table}[ht]
\caption{Multifractality Analysis of QKR: $\alpha$ values here are
obtained by an $L\rightarrow \infty$ extrapolation of $\alpha_L$
\cite{emm}. The $0 < D_2 <1$-behavior indicates a multifractal
nature of the three QKR cases.}
\centering
\begin{tabular}{c c c c}
\hline\hline
Case & \qquad QKR1 \qquad & \qquad QKR2 \qquad & \qquad QKR3 \\ [1.5ex]
\hline
$\alpha_0$ & 1.034 & 1.008 & 2.466\\
$\alpha_{1/2}$ & 1.007 & 1.001 & 0.812\\
$\alpha_1$ & 0.934 & 0.991 & 0.543\\
$D_0$ & 1 & 1 & 1 \\
$D_1$ & 0 & 0 & 0 \\
$D_2$ & 0.825 & 0.811 & 0.89 \\  [1.5ex]    
\hline
\end{tabular}
\label{table 1.}
\end{table}

\begin{figure}
\centering
\includegraphics[scale=0.85]{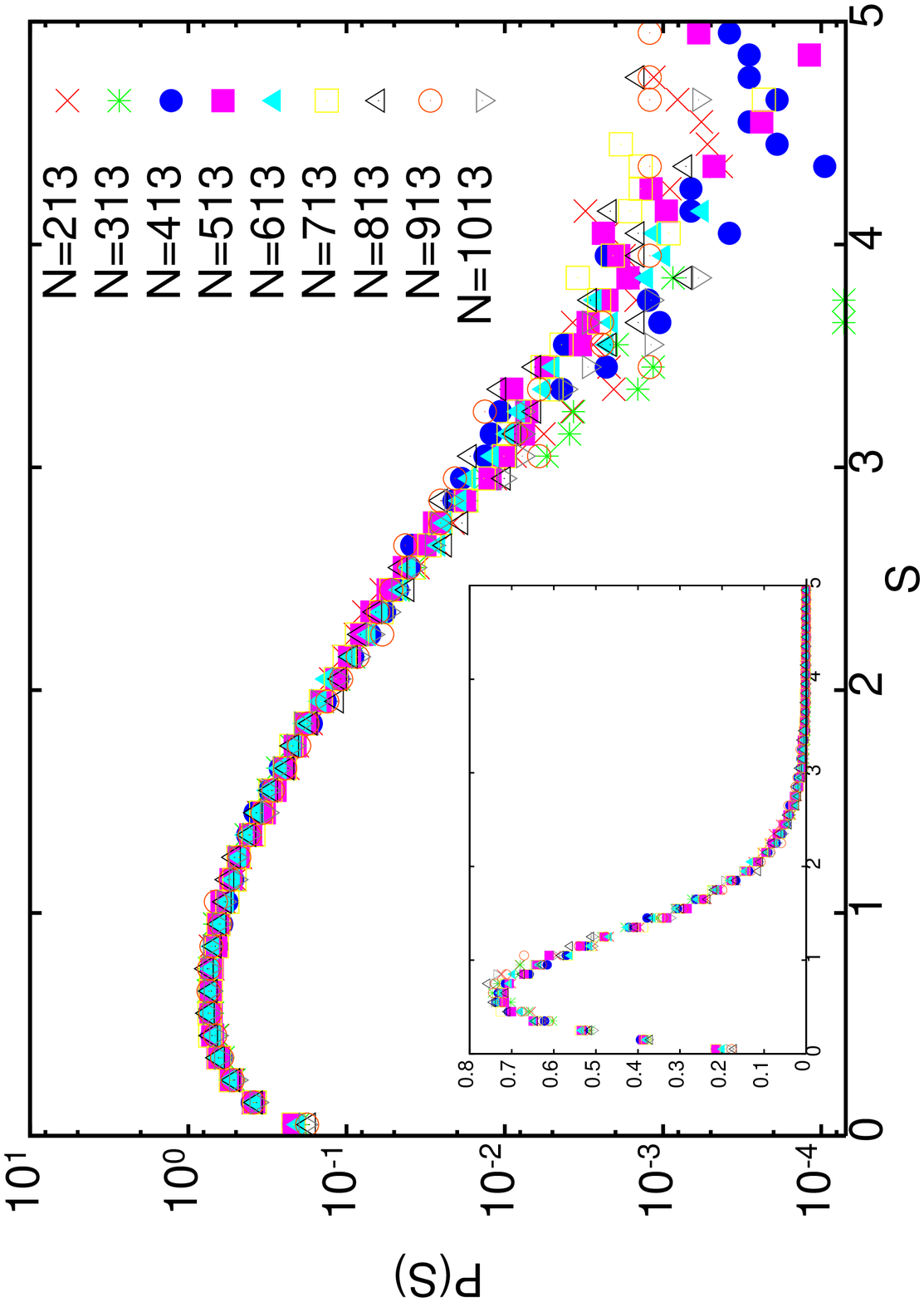}
\label{Fig.1(a)}
\end{figure}

\begin{figure}
\centering
\includegraphics[scale=0.85]{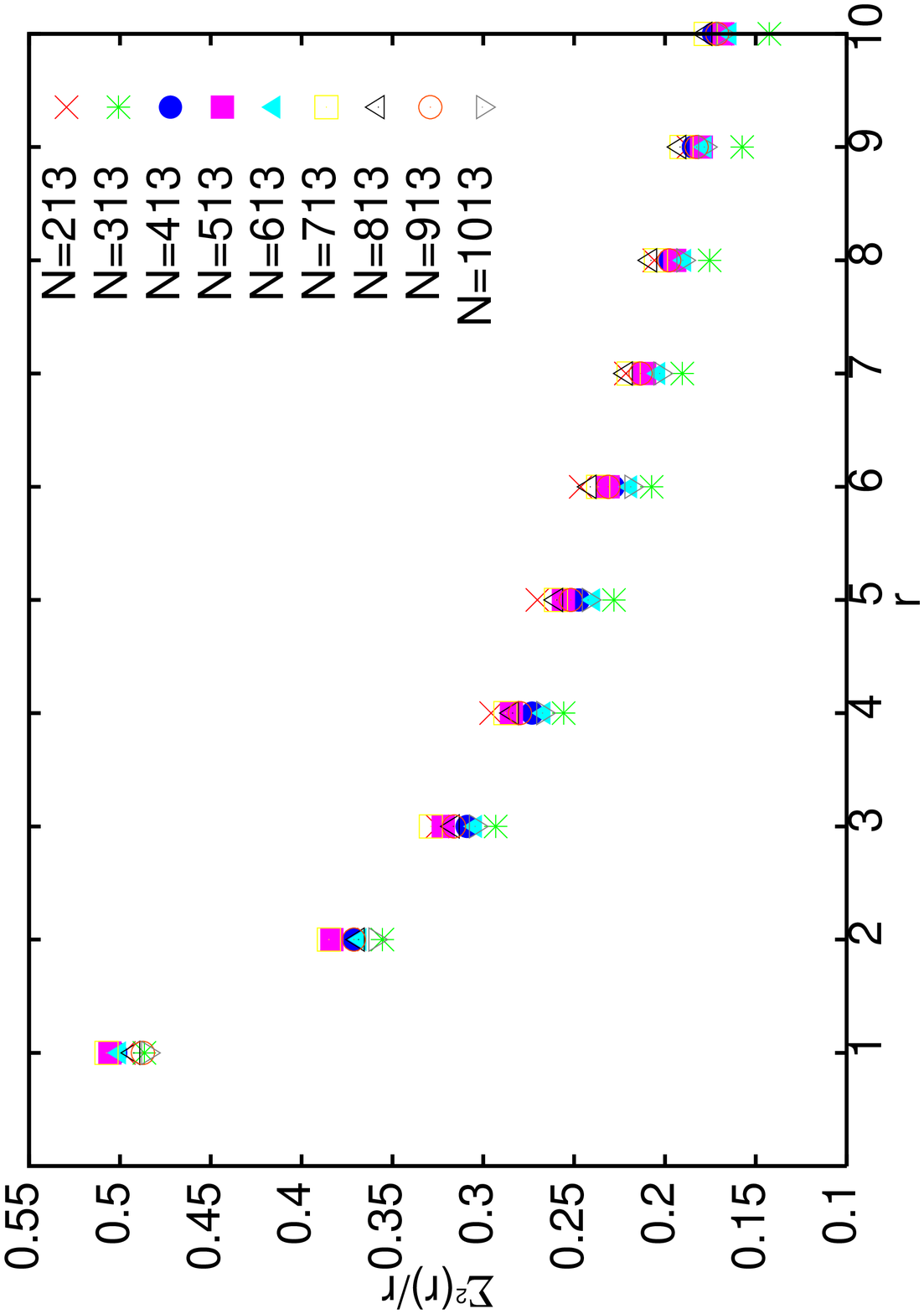}
\label{Fig.1(b)}
\end{figure}

\begin{figure}
\centering
\includegraphics[scale=0.85]{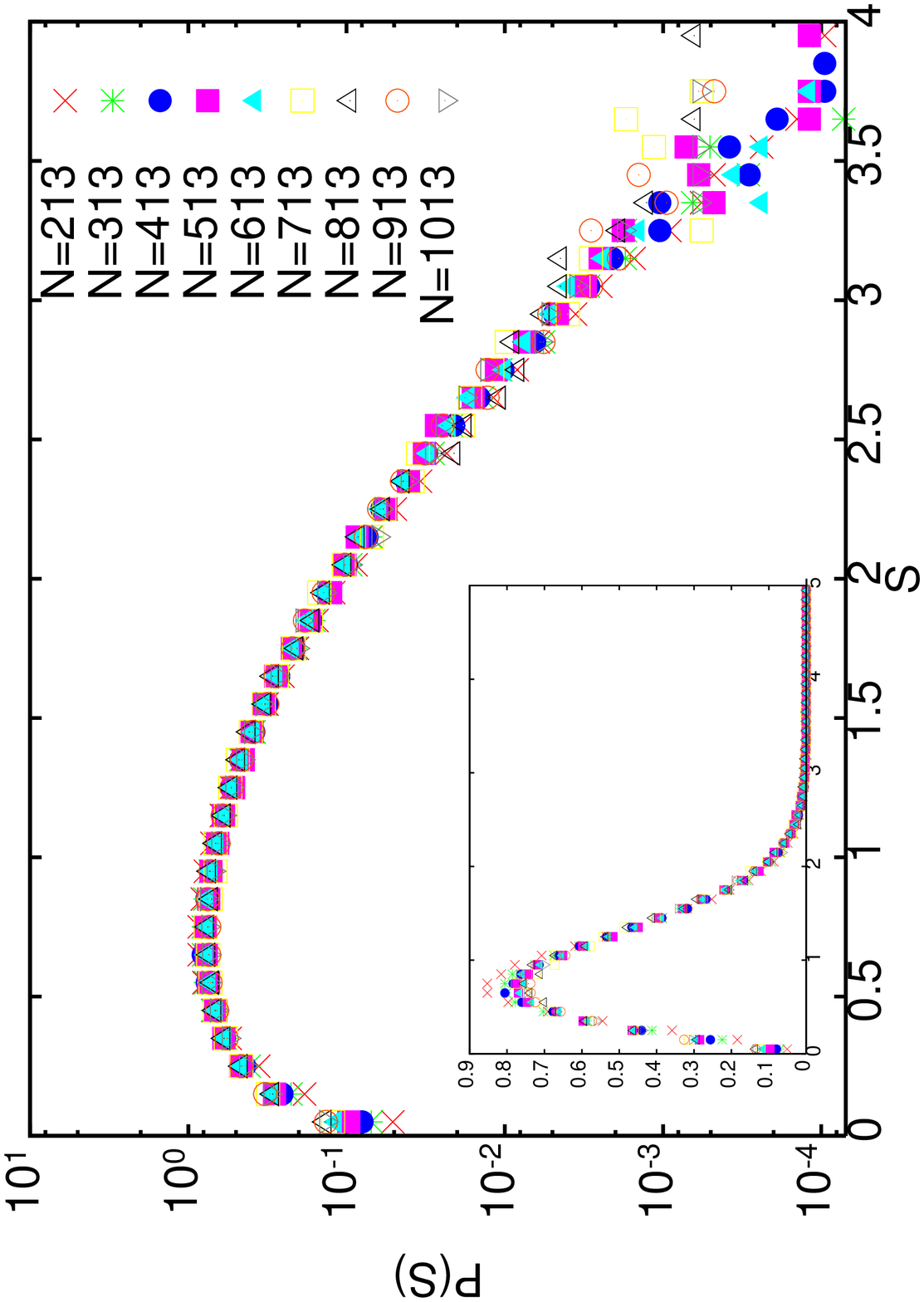}
\label{Fig.2(a)}
\end{figure}

\begin{figure}
\centering
\includegraphics[scale=0.85]{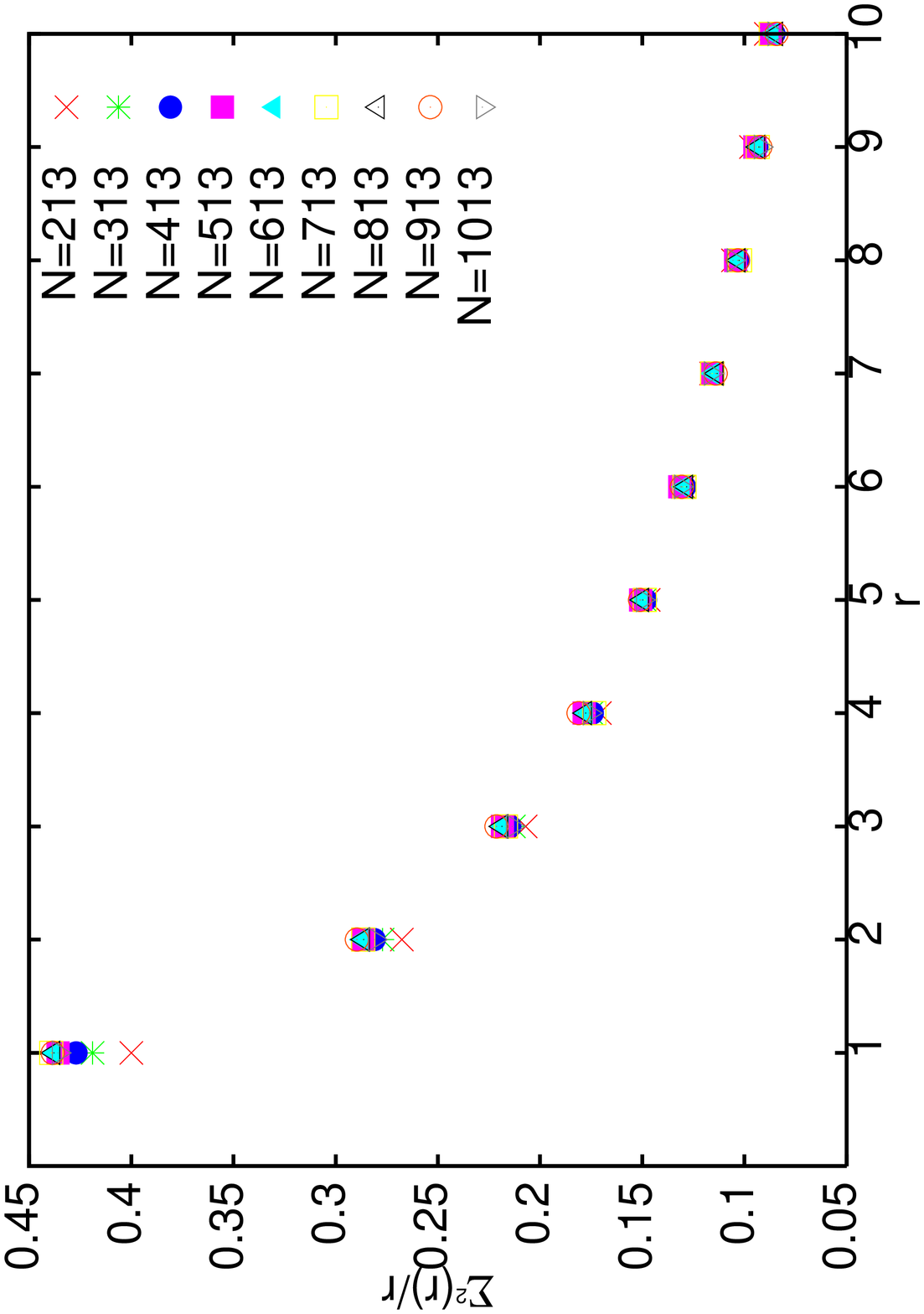}
\label{Fig.2(b)}
\end{figure}

\begin{figure}
\centering
\includegraphics[scale=0.85]{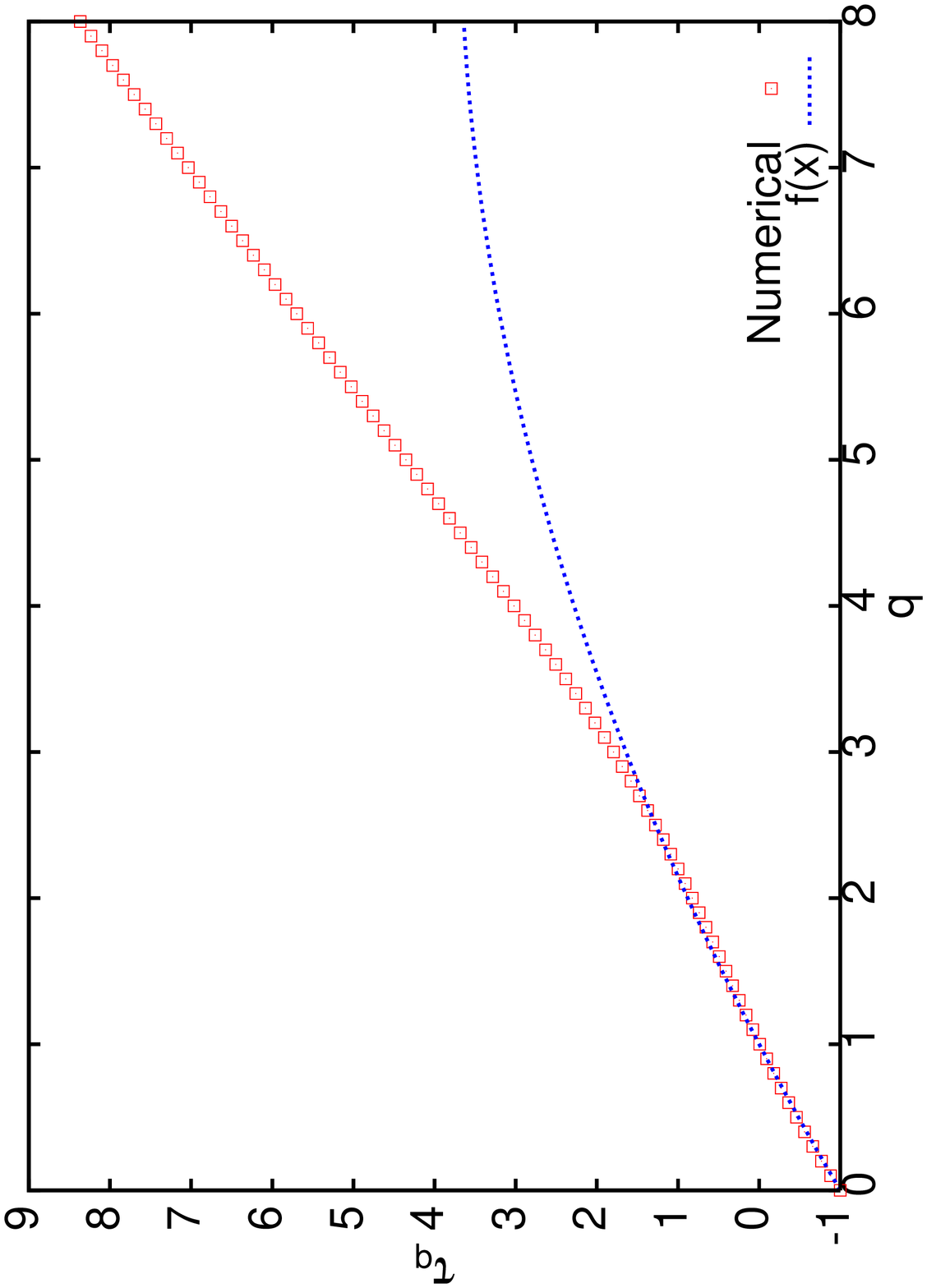}
\label{Fig.3(a)}
\end{figure}

\begin{figure}
\centering
\includegraphics[scale=0.85]{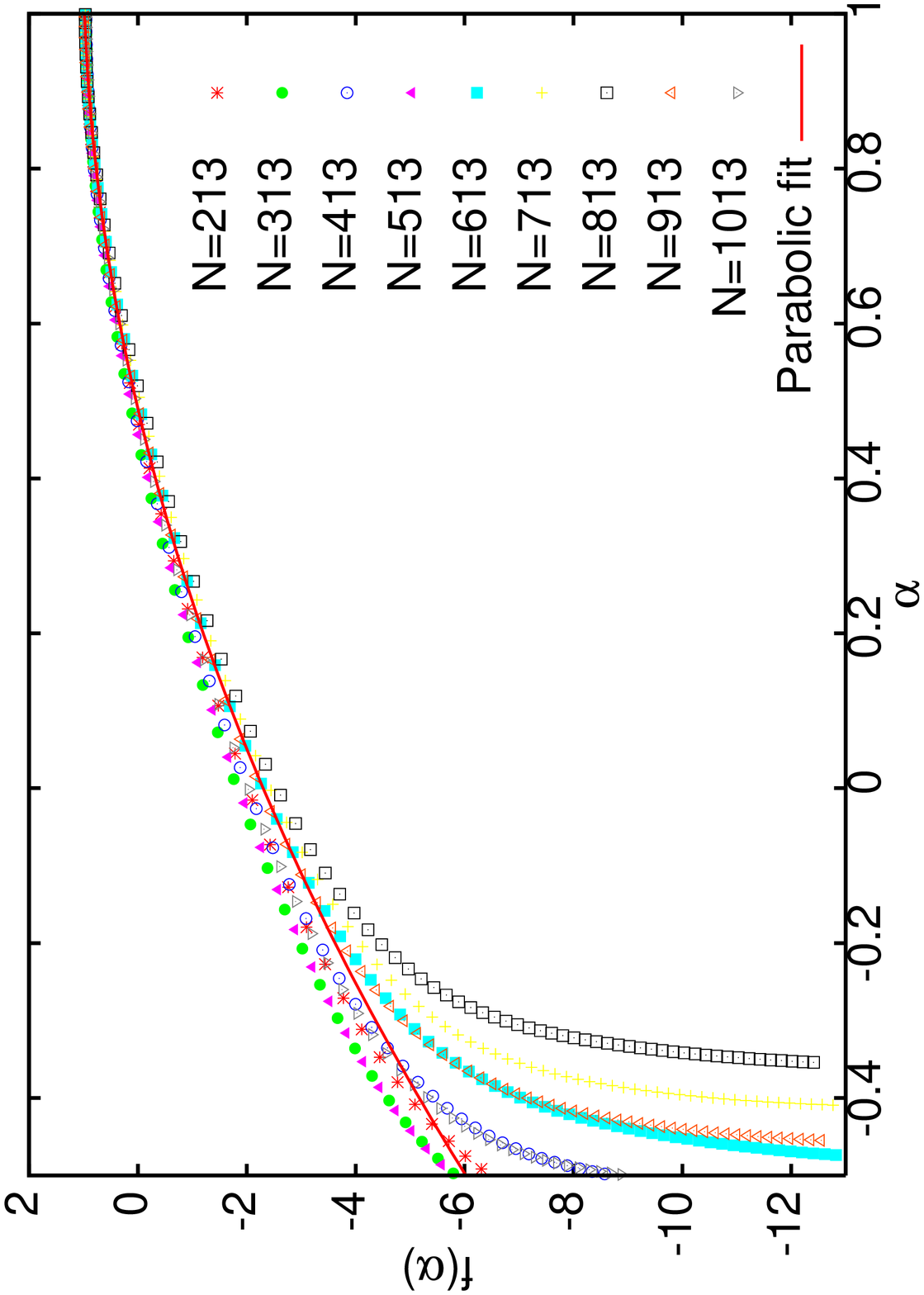}
\label{Fig.3(b)}
\end{figure}

\begin{figure}
\centering
\includegraphics[scale=0.85]{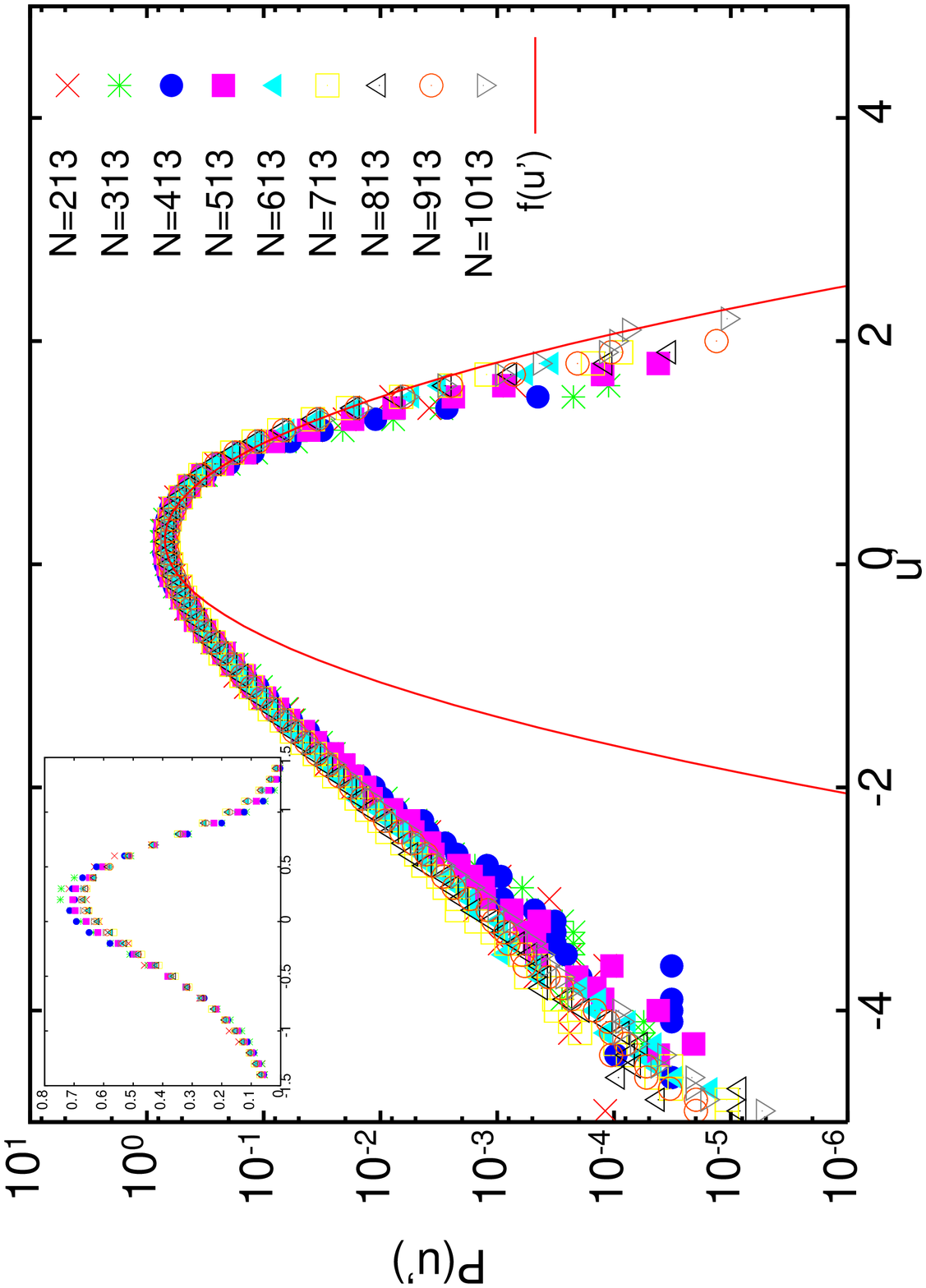}
\label{Fig.3(c)}
\end{figure}

\begin{figure}
\centering
\includegraphics[scale=0.85]{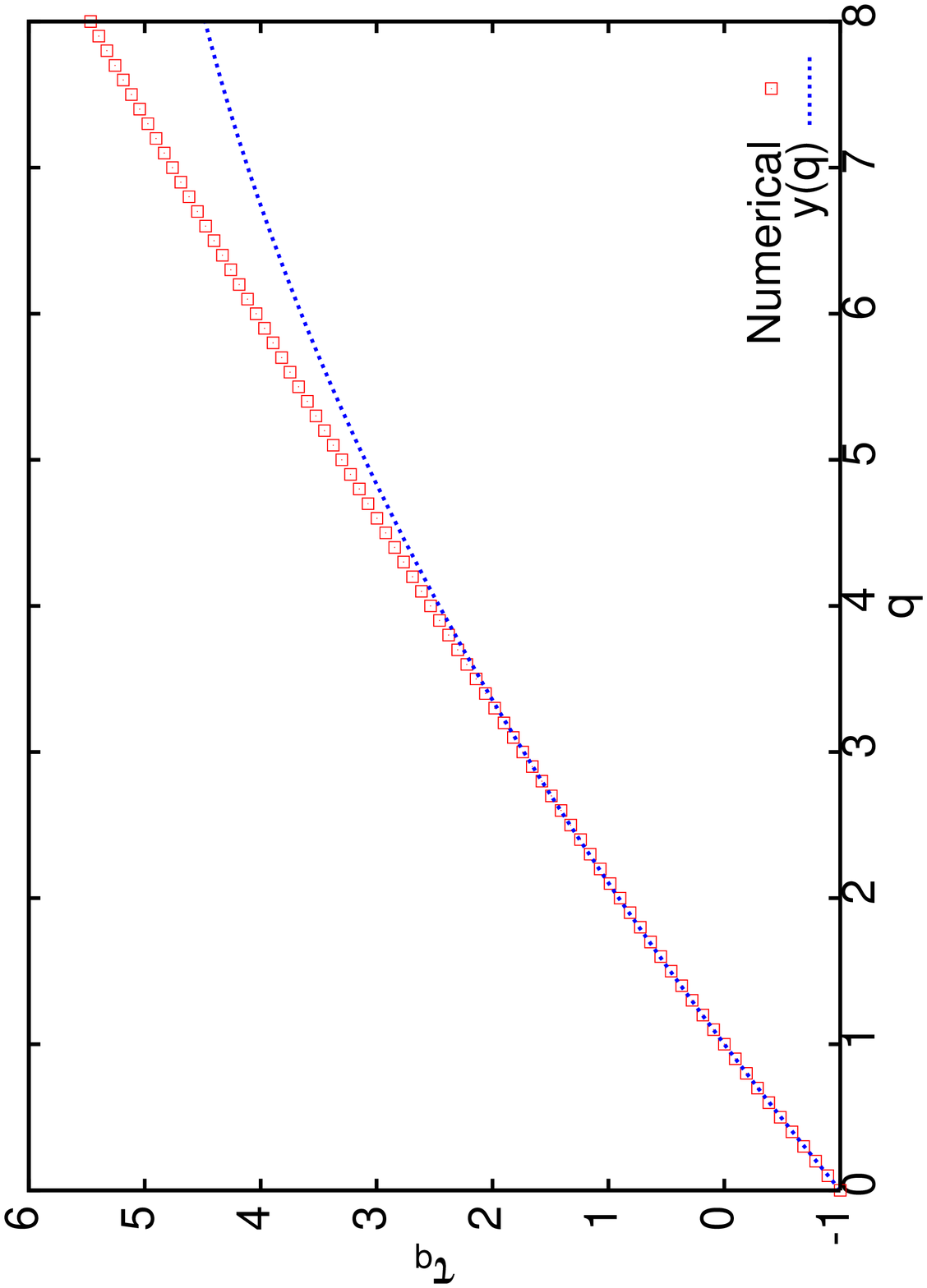}
\label{Fig.4(a)}
\end{figure}

\begin{figure}
\centering
\includegraphics[scale=0.85]{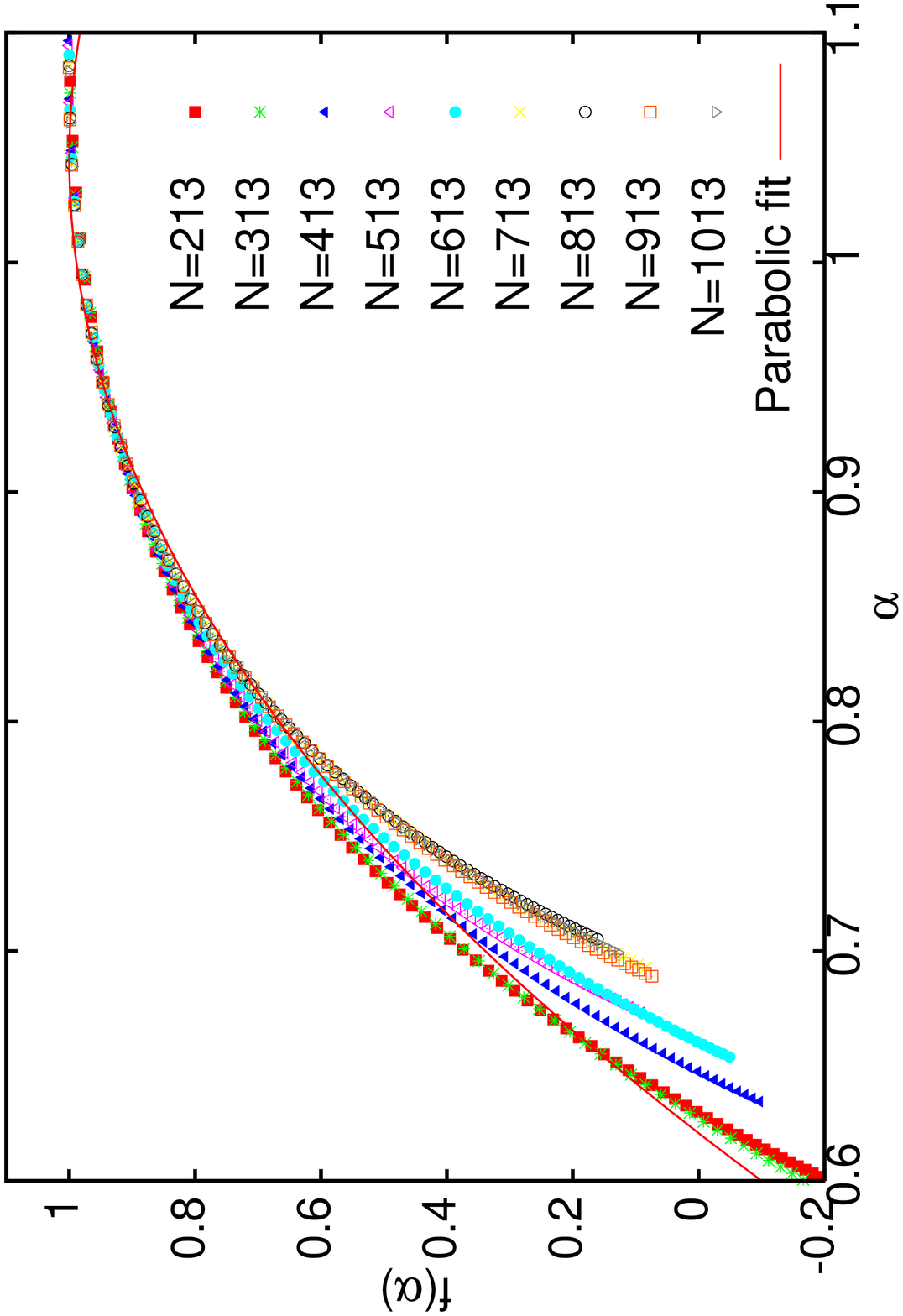}
\label{Fig.4(b)}
\end{figure}

\begin{figure}
\centering
\includegraphics[scale=0.85]{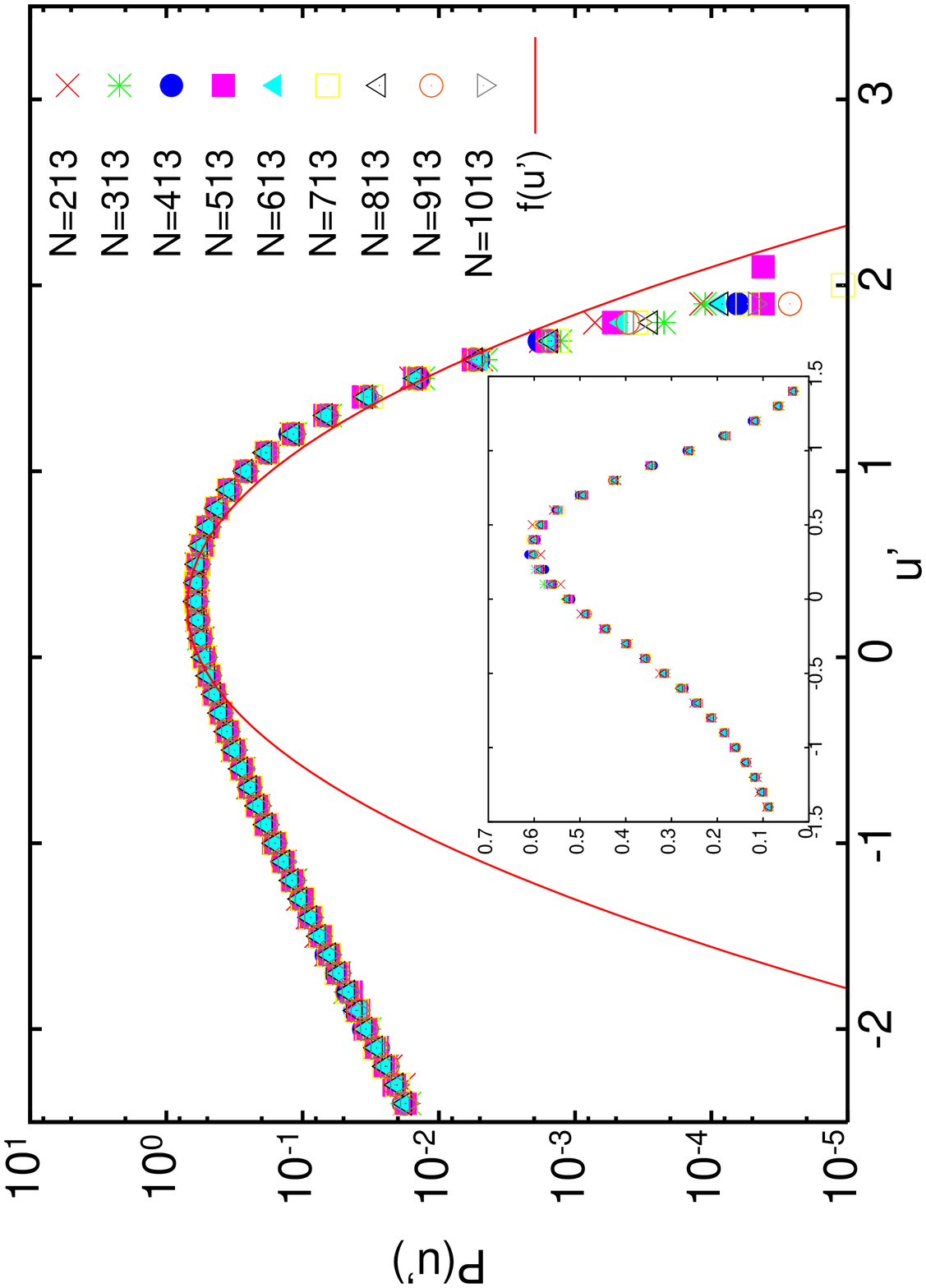}
\label{Fig.4(c)}
\end{figure}

\begin{figure}
\centering
\includegraphics[scale=0.85]{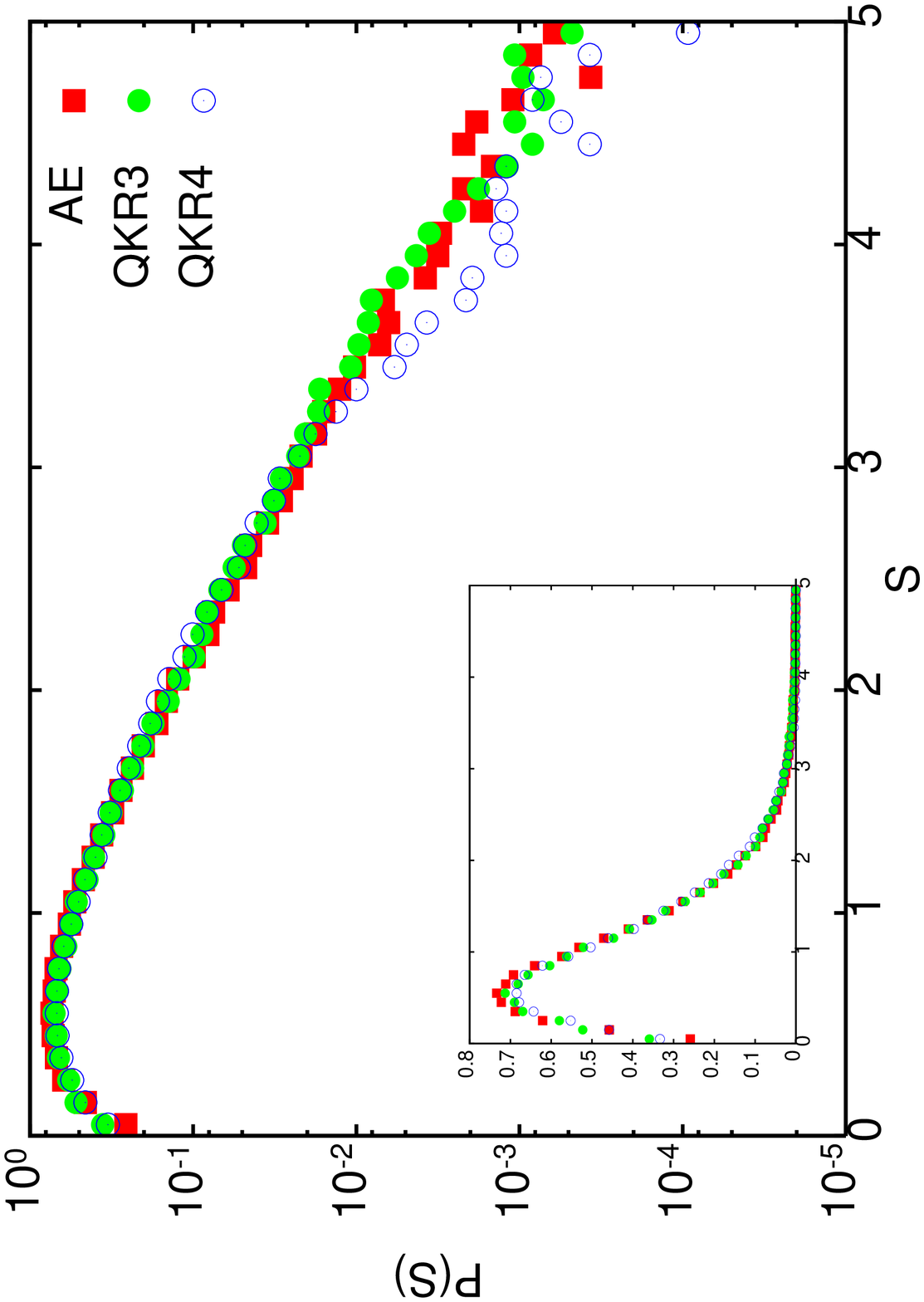}
\label{Fig.5(a)}
\end{figure}

\begin{figure}
\centering
\includegraphics[scale=0.85]{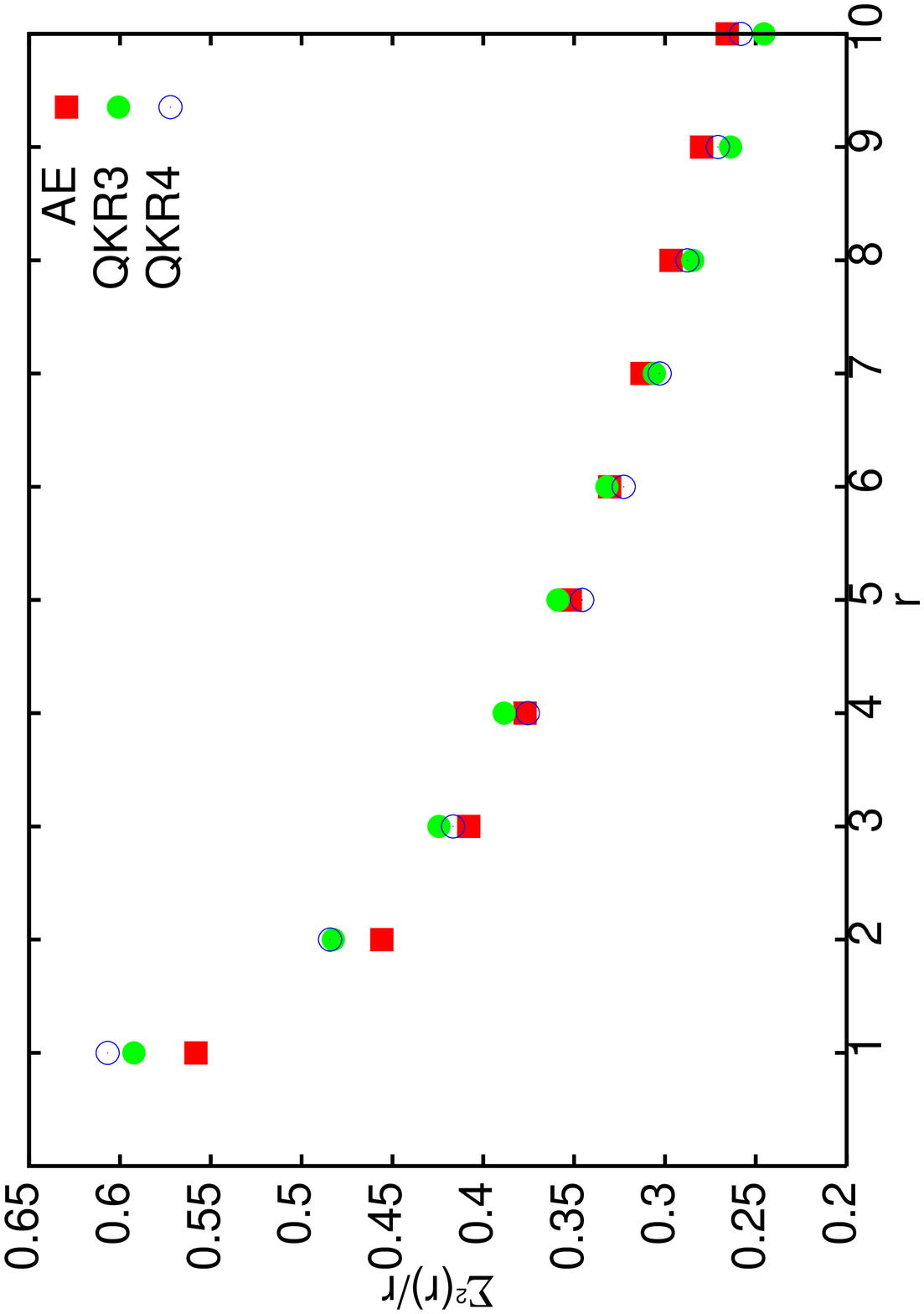}
\label{Fig.5(b)}
\end{figure}

\begin{figure}
\centering
\includegraphics[scale=0.85]{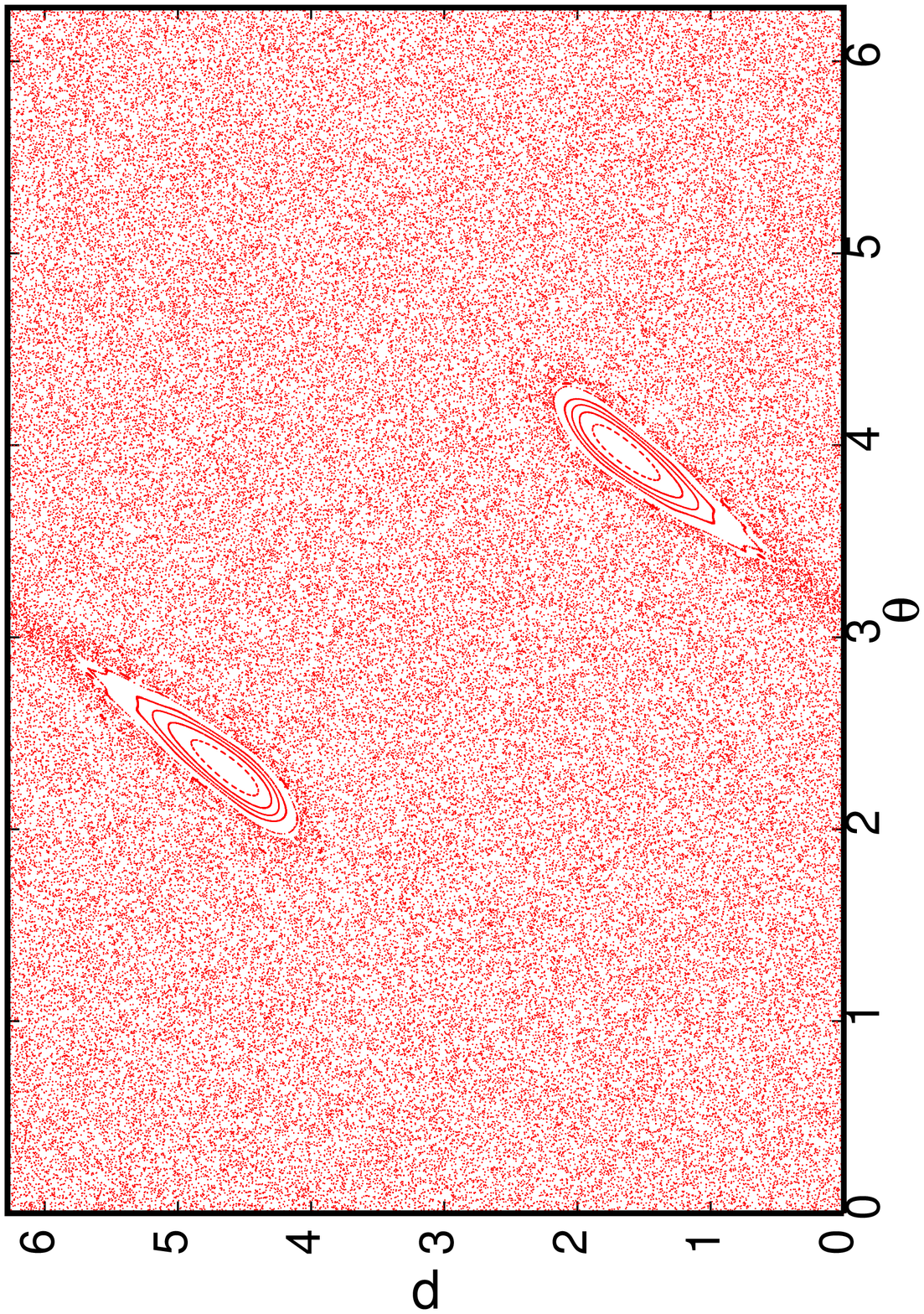}
\label{Fig.5(c)}
\end{figure}

\begin{figure}
\centering
\includegraphics[scale=0.85]{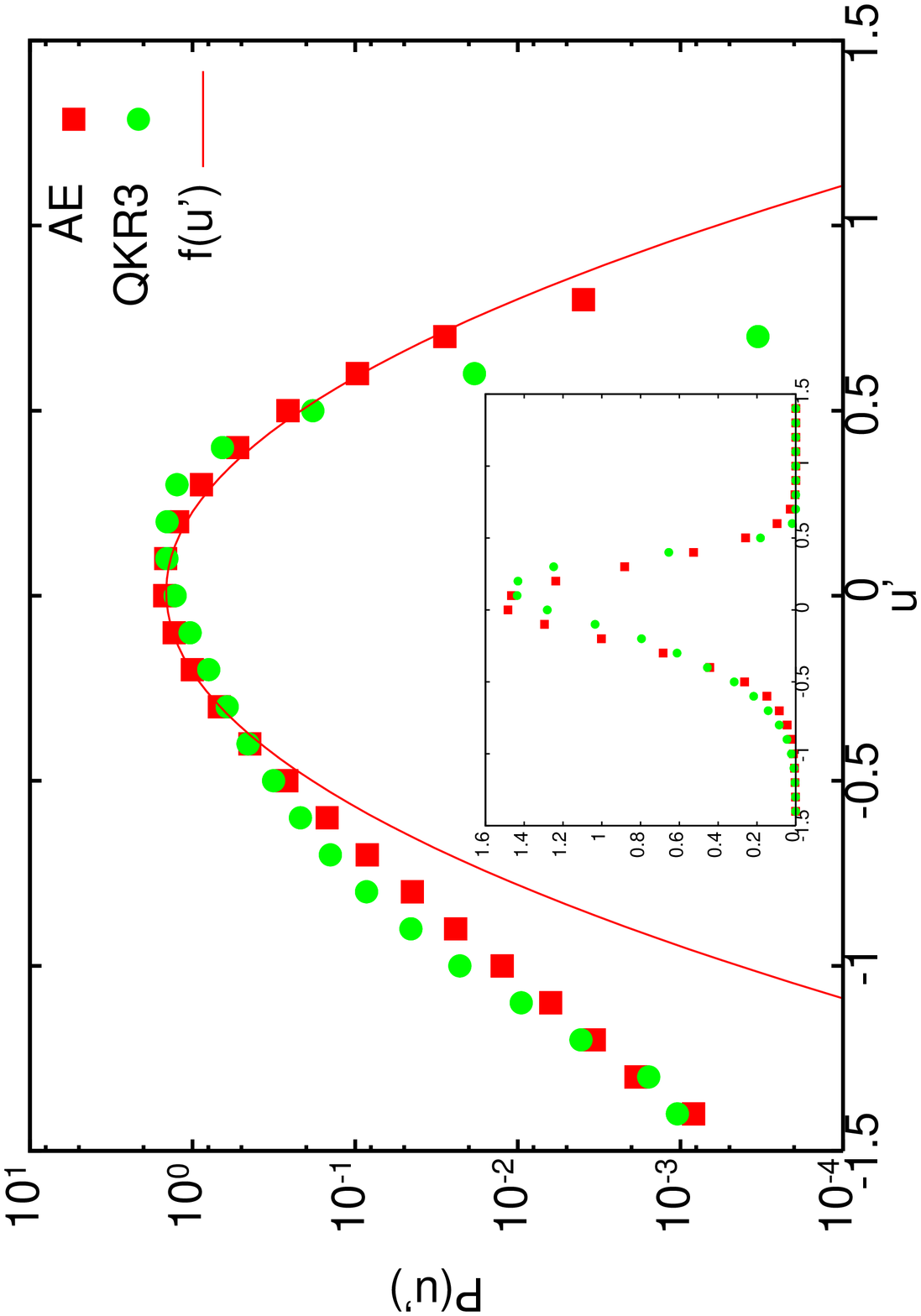}
\label{Fig.6(a)}
\end{figure}

\begin{figure}
\centering
\includegraphics[scale=0.85]{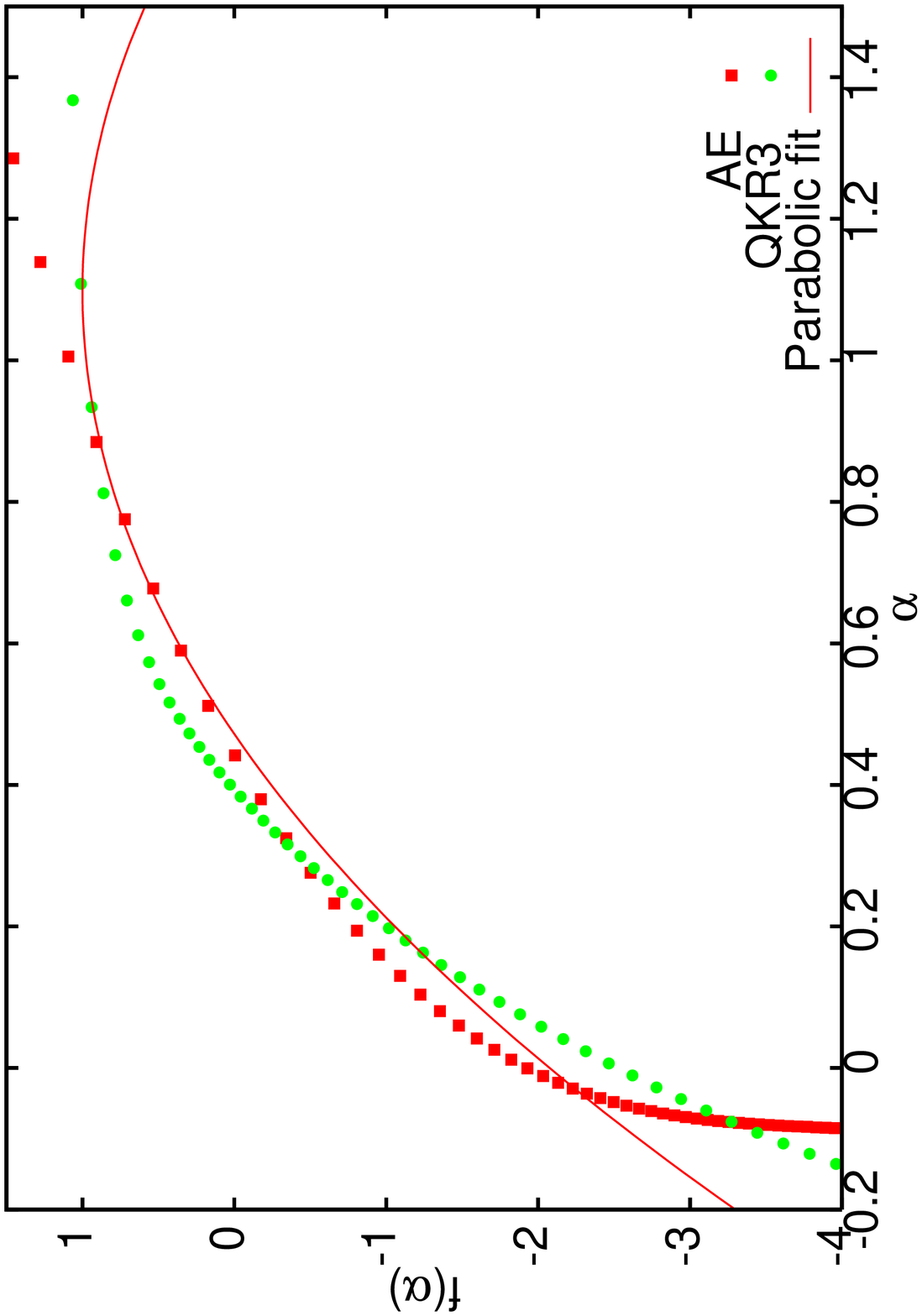}
\label{Fig.6(b)}
\end{figure}

\end{document}